\documentclass[journal]{IEEEtran}
\usepackage{cite}
\ifCLASSINFOpdf
\else
\fi
\usepackage[cmex10]{amsmath}
\usepackage{amssymb}
\usepackage{mathrsfs}
\usepackage{algorithm}
\usepackage{algorithmic}
\usepackage{multirow}
\usepackage{graphicx}
\usepackage{cite}
\usepackage{subfigure}
\usepackage{color}
\usepackage{amsmath}
\usepackage{amssymb}
\usepackage{amsmath}
\usepackage[top=2.5cm, bottom=2.5cm, left=2cm, right=2cm]{geometry}
\usepackage{flushend}
\usepackage{graphicx} %use graph format
\usepackage{epstopdf}

\begin{document}
\title{Content-Centric and Software-Defined Networking with Big Data}
\author{\IEEEauthorblockN{Haipeng Yao\IEEEauthorrefmark{1}, Chao Qiu\IEEEauthorrefmark{2}, Chao Fang\IEEEauthorrefmark{3},  Xu Chen\IEEEauthorrefmark{1}, and F. Richard Yu\IEEEauthorrefmark{4}}\\

%\IEEEauthorblockA{\IEEEauthorrefmark{1}
%Key Lab. of Universal Wireless Comm., Ministry of Education, Beijing University of Posts and Telecom., Beijing, P.R. China\\
%Email: zjkchouchao@bupt.edu.cn clzhao@bupt.edu.cn xufm@bupt.edu.cn}\\

\IEEEauthorblockA{\IEEEauthorrefmark{1}{\normalsize  State Key Lab. of Networking and Switching Tech., Beijing Univ. of Posts and Telecom.,  P.R. China}} \\

\IEEEauthorblockA{\IEEEauthorrefmark{2}{\normalsize Key Lab. of Universal Wireless Comm., Ministry of Education, Beijing Univ. of Posts and Telecom.,  P.R. China}} \\

\IEEEauthorblockA{\IEEEauthorrefmark{3}{\normalsize Beijing Advanced Innovation Center for Future Internet Technology and College of Information and Communication Engineering, Beijing University of Technology, Beijing, P.R. China}} \\

%\IEEEauthorblockA{\IEEEauthorrefmark{3}
%Beijing Advanced Innovation Center for Future Internet Technology, Beijing University of Technology, Beijing, P.R. China}}

\IEEEauthorblockA{\IEEEauthorrefmark{4}
Depart. of Systems and Computer Eng., Carleton Univ., Ottawa, ON, Canada}}

\maketitle

\begin{abstract}
Many communities have researched the application of novel network architectures such as Content-Centric Networking (CCN) and Software-Defined Networking (SDN) to build the future Internet. Another emerging technology which is big data analysis has also won lots of attentions from academia to industry. Many splendid researches have been done on CCN, SDN, and big data, which all have addressed separately in the traditional literature. In this paper, we propose a novel network paradigm to jointly consider CCN, SDN, and big data, and provide the architecture internal data flow, big data processing and use cases which indicate the benefits and applicability. Simulation results are exhibited to show the potential benefits relating to the proposed network paradigm. We refer to this novel paradigm as Data-Driven Networking (DDN).
\end{abstract}

\begin{IEEEkeywords}
Big data analysis, SDN, CCN, cache management, Data-Driven Networking.
\end{IEEEkeywords}

% \fi
\IEEEpeerreviewmaketitle

\section{Introduction\label{sect: introduction}}%% The very first letter is a 2 line initial drop letter followed
%% by the rest of the first word in caps.
%%
%% form to use if the first word consists of a single letter:
%% \IEEEPARstart{A}{demo} file is ....
%%
%% form to use if you need the single drop letter followed by
%% normal text (unknown if ever used by IEEE):
%% \IEEEPARstart{A}{}demo file is ....
%%
%% Some journals put the first two words in caps:
%% \IEEEPARstart{T}{his demo} file is ....
%%
%% Here we have the typical use of a "T" for an initial drop letter
%% and "HIS" in caps to complete the first word.
\IEEEPARstart{T}{he} current Internet architecture established from TCP/IP has gained huge {\color{black}success} and been the one of indispensable infrastructures for our daily life, economic operation and society. However, burgeoning mega trends in the information and communication technology (ICT) domain are urging the Internet for pervasive accessibility, broadband connection and flexible management, which call for potential new Internet architectures. The original design tactic of Internet, which is ``Leaving the complexity to hosts while maintaining the simplicity of network" \cite{hu_survey_2011}, leads to the almost insurmountable challenge known as ``Internet ossification": software in the application layer has developed rapidly, and abilities in the application layer have been drastically enriched. By contrast, protocols in the network layer lack scalability and the core architecture is hard to modify, which {\color{black}means} that new functions have to be implemented through myopic and clumsy ad hoc patches in the existing architecture. For example, the transition from IPv4 to IPv6 is difficult to deploy in practice \cite{wu_transition_2013}.

To improve the performance of the current Internet, novel network architectures have been proposed by the research communities to build the future Internet, such as Content-Centric Networking (CCN) and Software-Defined Networking (SDN). CCN is {\color{black}a} rising networking paradigm centered on content distribution rather than host-centric connectivity \cite{perino_reality_2011}. SDN is another paradigm, which separates the control plane from forwarding plane, breaks vertical integration, and introduces the ability of programming the network \cite{sdnopenissue,YYG15,YY15,CYH16,HYH16}.

In addition, big data has won great attentions in terms of academia and industry. Big data represents the data sets are so large and complex that traditional data management tools or processing methods are inadequate to manage and analyze them. Big data is popularly characterized by ``5Vs" (initially it was described as ``3Vs", and two have been added recently): Volume (the size of data sets), Variety (the range of data types and sources), Velocity (the speed of data in and out), Value (how useful the data is), and Veracity (the quality of data) \cite{GBR15}. It could bring lots of advantages to networking, such as management (recognize-requirement) and intelligence (recognize-variation), and has the potential to make sure that we operate and manage data networks. {\color{black}In this regard, some excellent works have explored the role of big data in the traditional networks. The authors of \cite{zhang_social_2016,HYZ16} introduce and explore the features and categories of mobile data, which is extremely beneficial to wireless networks. On this basis, some big-data-enabled architectures in wireless network are proposed in \cite{zeydan_big_2016} and \cite{bastug_big_2015}. In other networks, big data also plays a key role, such as \cite{YL01,MYL04,CYY15,XYJL12,LY15,BYC12,wang_mobile_2016,YWL06_MONET,YTM10,HYZ16,ZYN12_JSAC,YK07,jiang_energy_2016,WTY14,LYH10,lin_system_2016}. }

Albeit many splendid researches have been done on CCN, SDN, and {\color{black}big data for networks,} they all have been addressed separately in the traditional literature. Nevertheless, as listed in the following, it is necessary to consider them together to offer better services in the future network.
\begin{itemize}
    \item Firstly, CCN has been considered as the one of promising architectures for efficient content distribution around the Internet.  This paradigm shifting from host-centric to content-centric has many alluring advantages, {\color{black}for example} the reduction of network load, low latency and so on \cite{LYZ15,WYL16,LYY16,FYH15,WLY16}. Currently, there are increasing number of researches in this field, such as NDN \cite{zhang_named_2010}, PURSUIT \cite{fotiou_developing_2012}, SAIL \cite{leva_description_2011} and so on. The major feature of CCN is in-network caching \cite{fang_survey_2014}. This good feature has significant impacts on providing content to users in SDN. In this paper, we propose to add caching capacity in SDN switches, which enables to cache content so as to reduce the content response time and provide improved users experiences. When a certain content is sent to reply a user's request, this content will be cached in SDN switches along the way back to this request originator. With the in-network caching capacity, the performance of SDN can be improved in terms of content distribution latency.
    \item Secondly, SDN can contribute to the promotion of CCN. One of biggest challenges for global optimal network and content cache management of CCN are inherently distributed, where every node has only a partial view. In particular, SDN has centralized control plane, decouples control from data plane, which will help CCN allocate cache resources, distribute {\color{black}content}, and configure networks globally. For example, by knowing that a switch needs more cache resources, the control layer can send flow tables to the infrastructure layer to allocate more caches to this switch.
    \item Finally, big data has profound {\color{black}impact} on the design and operation of SDN. Particularly, with the global view of the network, the logically centralized controller in SDN can obtain big data from all the different layers (i.e., from infrastructure layer to application layer) with arbitrary granularity. Using big data analysis in the control layer of SDN, we can extract knowledge out of the large volumes of data to help the controller make decisions. For example, with big data analysis, we {\color{black}will} know which content in a certain switch has the high popularity. Based on the analysis results, the control layer enables to redistribute content, which makes the required {\color{black}content} close to {\color{black}the} specific users.

\end{itemize}

For the above reasons, we adopt the centralized control by SDN, combined with distributed in-networking caching provided by CCN. In the context, big data analysis can extract knowledge about network, and send the knowledge to control the whole network by centralized control capacities offered by SDN. We refer to this new paradigm gathering with SDN, CCN and big data analysis as Data-Drive Networking (DDN).

The rest of this paper is organized as follows. We describe the Data-Driven Networking (DDN) paradigm and how it operates in Section~\ref{sect: architecture}. In Section~\ref{sect: dataflow}, data flow in Data-Driven Networking is discussed. Section~\ref{sect: processing} shows the data processing technology. Then we describe relevant use cases to show the validity and applicability of DDN paradigm in Section~\ref{sec:usecases}. At last, we conclude the study in Section~\ref{sect: con}.

\section{Data-Driven Networking architecture\label{sect: architecture}}
\begin{figure}[tp]
\centering
\centering\includegraphics[width=8.75cm]{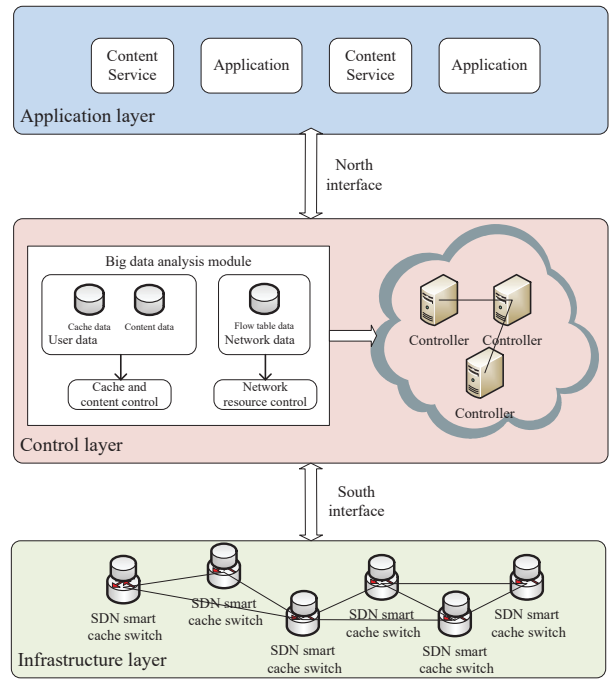}
\caption{Architecture reference model: a three layer model, ranging from the infrastructure layer to the control layer to the application layer.}
\label{fig: fig1}
\end{figure}

The reference model for the Data-Driven Networking is shown in Fig.~\ref{fig: fig1}. This model consists of three layers, namely the infrastructure layer, the control layer and the application layer, with two interfaces (i.e., south interface and north interface) in a bottom-up manner.

The \emph{infrastructure layer} is responsible for storing {\color{black}content}, monitoring and forwarding data packets. We embed content caches in the traditional SDN switches to provide {\color{black}content} for users, which achieves significant reduction in content response time and provides improved users' experiences. When users' requests arrive at a switch, this switch firstly finds the needed {\color{black}content} in its own cache. If the needed {\color{black}content is} in this switch cache, it will quickly respond to users' requests. Otherwise, the switch will retrieve {\color{black}content} from the source by packet forwarding. Besides, there are monitor agents in content caches. They are responsible for collecting content data, cache data, and network data, and sending them to the big data analysis module in control layer through the south interface. Finally, the same as traditional data plane in SDN, switches are composed of forwarding hardware, operate unaware of the network according to flow tables and update the configuration. And they are enable to adjust the size and {\color{black}content} of cache based on instructions from the control layer, which ensures the reasonable cache reallocation and content redistribution.

The \emph{control layer} connects the infrastructure layer and the content service layer, via the two interfaces. The control layer is the most significant core in this architecture, in which the complexity resides. It consists of the below two aspects, namely big data analysis module and ordinary SDN controller module.

The \emph{application layer} includes diverse application services to satisfy users' requirements. Taking advantage of north interface and control layer, applications enable to access the global network view and programmatically implement strategies to leverage the physical networks at the infrastructure layer using the high level language.

{\color{black}The cooperation with each components will be described in the following section.}

\section{data flow in data-driven networking\label{sect: dataflow}}
\begin{figure}[tp]
\centering
\centering\includegraphics[width=8cm]{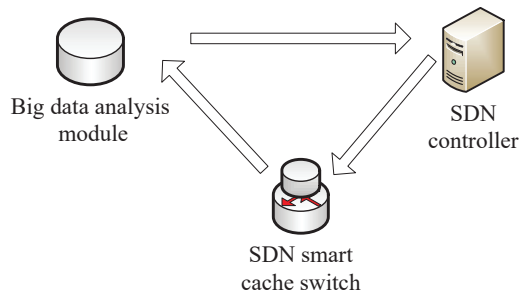}
\caption{DDN data flow}
\label{fig: fig2}
\end{figure}

Data-Driven Networking paradigm runs under the data flow. {\color{black}This is the brightest different between existing works which have been done combing CCN or ICN with SDN, namely the flows from SDN smart cache switch to big data analysis module, from big data analysis module to SDN controller, and from SDN controller to SDN smart cache switch.} Fig.~\ref{fig: fig2} only indicates the emerging data flow in DDN system. Some original data flows in traditional SDN such as from data plane to control plane are not listed. In what follows we describe these flows in detail.
\subsection{From SDN smart cache switch to Big data analysis module}

Big data analysis module aims to collect enough data with arbitrary granularity to complete the view of network. Therefore, monitor agents in switches gather the content data, cache data, and network data, send them to big data analysis in control plane through the south interface in real time when switches forward packets. The most related data gathered by switches is as follows.

\begin{itemize}
    \item User data: user data consists of cache data, content data and history request data from SDN smart cache switches. In detail, cache data includes the total cache size and remaining cache size in each cache. Content data covers content types (e.g., text, picture or  video), specific content information (e.g, learning text about mathematics, landscape painting about Big Ben, or movie about Interstellar), bandwidth used for transferring this content, and content requirements for time delay and packet loss rate. Meanwhile, history request data is mainly about the request times of each user aiming at each content. User data {\color{black}is described} in Table~\ref{Tab: userdata}.
        \begin{table}[!hbp]
        \centering
        \caption{User data}
        \label{Tab: userdata}
        \begin{tabular}{|c|c|}
        \hline
        User data  & Remarks \\ \hline
        \multirow{2}{*}{Cache data}&
        Total cache size \\
        \cline{2-2}
        &Remaining cache size\\
        \hline
        \multirow{2}{*}{Content data}&
        Content type\\
        \cline{2-2}
        & Content information\\
        \cline{2-2}
        &Bandwidth\\
        \cline{2-2}
        &Time delay\\
        \cline{2-2}
        &Packet loss rate\\
        \hline
        History request data & Request times\\
        \hline
        \end{tabular}
        \end{table}

    \item Network data: it includes the physical, topological states of network, especially flow tables. A flow table consists of header field, counter and actions.The counter supports data statistics at four granularities: flow table, flow, port and queue. It is used to count traffic information, such as active entries, packet lookups and received bytes.The statistics in flow table is shown in Table~\ref{Tab:flowtable}.
    \begin{table}[!hbp]
    \centering
    \caption{The statistics in Flow table}
    \label{Tab:flowtable}
    \begin{tabular}{|c|c|}
    \hline
    Counter  & Bits \\ \hline
    \multicolumn{2}{|c|}{Per Flow table}\\
    \hline
    Active entries &32\\
    \hline
    Packet lookups &64\\
    \hline
    Packet matches &64\\
    \hline
    \multicolumn{2}{|c|}{Per Flow}\\
    \hline
    Received packets &64\\
    \hline
    Received bytes&64\\
    \hline
    Duration/seconds &32\\
    \hline
    Duration/nanoseconds &32\\
    \hline
    \multicolumn{2}{|c|}{Per Port}\\
    \hline
    Received packets &64\\
    \hline
    Transmitted packets &64\\
    \hline
    Received bytes &64\\
    \hline
    Transmitted bytes &64\\
    \hline
    Receive drops &64\\
    \hline
    Transmit drops &64\\
    \hline
    Receive errors &64\\
    \hline
    Transmit errors &64\\
    \hline
    Receive frame alignment errors &64\\
    \hline
    Receive overrun errors &64\\
    \hline
    Receive CRC errors &64\\
    \hline
    collisions &64\\
    \hline
    \multicolumn{2}{|c|}{Per Queue}\\
    \hline
    Transmit packets &64\\
    \hline
    Transmit bytes &64\\
    \hline
    Transmit overrun errors &64\\
    \hline
    \end{tabular}
    \end{table}
    \end{itemize}

The volume of above data is terribly huge. According to the latest data released by the International Telecommunication Union (ITU), the number of global Internet users was 3.2 billion by the end of 2015, accounting for almost 40 percent of the world's population. In the SDN paradigm, the data related to control information of users is greatly enormous. Meanwhile, the {\color{black}content of the Internet is more abundant}. In September 2014, the number of global websites was more than 1.06 billion, and this number is {\color{black} still} growing. In China, as of June 2015, according to the data from the 36th China Internet Network Development State Statistic Report,  the number of websites was 3.57 million. The average number of webpages of a website was 46,900, with the increase of 2.3 percent, and the number of bytes per webpage was 50KB, with the increase of 19 percent, which show that the {\color{black}content of the Internet is greatly rich}. Briefly, user data and network data are tremendous, and the network needs big data analysis.

The big data analysis module structures a collective intelligence data architecture with built-in analytics capabilities and its detailed processing procedure will be presented in Section~\ref{sect: processing}.  After gathering data from switches, big data analysis module extracts knowledge out of the large volumes of data, which has influence on building traffic model, users' behavior model and content popularity model so as to comprehend network status, users' demand and heat of content. Based on the cache data, we use big data analysis to know which switch has a higher frequency of content request in order to redress cache allocation. Based on the content data, we utilize big data analysis to know which content in a certain switch has a higher  demand in order to readjust content distribution. Based on the network status and big data analysis, we know which links are easier to congest or which switches are more possible to loss packets in order to make reasonable forwarding decisions.
\subsection{From Big data analysis module to SDN controller}
After making decisions such as cache reallocation, content redistribution and network management policies, big data analysis modules send them to SDN controller. Then SDN controller translates these policies into specific control actions. Because SDN controller has the global network view and manages the whole devices in data plane uniformly. In this sense, SDN controller enables to form the accurate control actions to execute analytical results. Finally, SDN controller packages actions into {\color{black}flow tables}.
\subsection{From SDN controller to SDN smart cache switch}
SDN controller sends flow tables to the switches through {\color{black} the southbound interface} to configure the infrastructure layer on the basis of decisions made by big data analysis. In detail, switches' behavior and caches' behavior will be changed, including packet forwarding rules, cache allocation and content distribution, in order to achieve maximum link and cache resource utilization, optimize content distribution and minimize link congestion.
\section{Big Data Processing Platform\label{sect: processing}}
\begin{figure}[tp]
\centering
\centering\includegraphics[width=8cm]{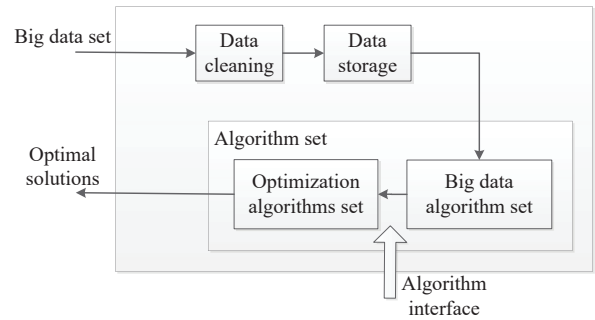}
\caption{The big data processing platform.}
\label{fig: fig3}
\end{figure}

The big data processing platform in this architecture consists of the following three parts: data cleaning, data storage, algorithm set that includes big data algorithm set and optimization algorithm set. This big data processing platform is shown in Fig.~\ref{fig: fig3}.

\begin{figure}[tp]
\centering
\includegraphics[width=8cm]{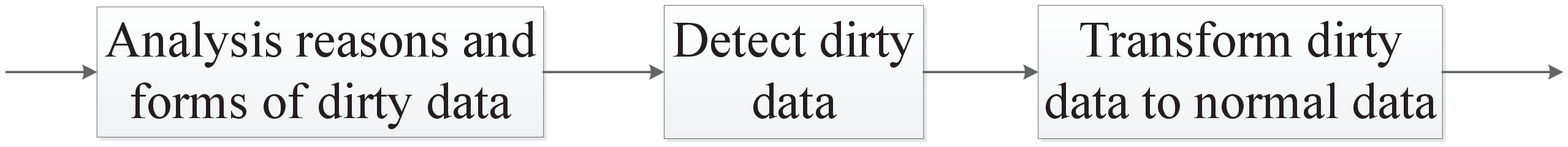}
\caption{The procedure of data cleaning.}
\label{fig: fig4}
\end{figure}

Firstly, the multi-source data comes from different traffics, different hardware platforms or different operating systems and so on. Inevitably, qualities issues of data  exist in the system. For example, similar or repetitive data abnormal data and incomplete data, which all can be called as ``dirty data". And data cleaning is the final mechanism that enables to find and correct these dirty data in data files. The procedure of data cleaning is shown in Fig.~\ref{fig: fig4}. Based on analytical results of causes and existing forms of dirty data, we detect dirty data in the system by technical methods. Finally the detected dirty data is transformed into normal data that meets the requirements of data quality. The ideology of data cleaning is backtracking, which analyzes data set from data source and detects every places of data set going through so as to extract data cleaning algorithms, rules and polices. Finally, these algorithms, rules and polices use  the data set to find dirty data and clean dirty data. After data cleaning, the accuracy of data can be improved.

Secondly, after data cleaning, the data set is stored for further processing. Since the data size in this architecture is greatly huge, SQL is not suitable. We consider NoSQL as a better choice in this architecture. NoSQL is shortened from Not Only SQL, which adopts flexible, distributed, extensible data storage management to meet the needs of big data storage. According to the classification of storage model, NoSQL consists of extended column storage, such as Google BigTable, HBase, Cassandra, Key-Value storage, such as Redis, BerkerlyDB, graph storage, such as Neo4J, Infinite Graph, document storage, such as MongoDB, CouchDB, and so on.

Thirdly, there are lots of big data algorithms in big data algorithm set, such as C4.5, K-means, Support vector machines(SVM), Apriori, EM, PageRank, AdaBoost, KNN, Naive Bayes, CART and so on. There are also lots of optimization algorithms in the optimization algorithm set, such as conjugate gradient algorithm, steepest descent algorithm, penalty function algorithm, simplex algorithm, newton algorithm, levenberg-marquardt algorithm, variable elimination algorithm, and gradient projection algorithm. Meanwhile, the algorithm set has the external interface by which new algorithm can be added in the set. Different types of data enter the algorithm set. According to its own characteristics, different data chooses the appropriate big data algorithm in the algorithm set. According to different optimization objectives, different data chooses the appropriate optimization algorithm in the algorithm set. Through the big data algorithm and optimization algorithm, aiming at some optimization objectives, we are enable to get the optimal solutions.

\section{use cases\label{sec:usecases}}
In this section, we present the specific use case which explains the workflow and specific example of DDN. We also give simulation performance of DDN paradigm.
\subsection{Workflow}
A user submits a content request to its nearest SDN smart cache switch. If the {\color{black}content is} cached in this switch or this switch knows how to obtain the {\color{black}content} from other switches based on the flow tables, this switch will reply user's request immediately or get the required {\color{black}content} based on the flow table to reply user's request. Otherwise, this switch will upload the request to the controller, which asks for the flow tables to reply user's request. {\color{black}This process is just the same as traditional SDN and there is no need for big data analysis module to make decisions aiming st this flow. Meanwhile, the monitor in this switch records request frequency of this content and whether this request is hit. At stated times,} the big data analysis module gathers data from the infrastructure layer, comprehends network status, performance and users behavior, and sends the results of learning to the controller, which will have positive impacts on entire network management, cache resource management and content distribution management. For example, a certain user sends a content request of film ``Avatar" to its nearest SDN smart cache switch. Unfortunately, Avatar is not cached in this switch, and this switch doesn't know how to get Avatar based on the flow tables. So this switch uploads this request to the controller to ask for the flow tables, and gets Avatar based on the flow tables. Meanwhile, after a period of time, the big data analysis module learns that the users connected to this switch usually ask for Avatar, and tells this result of learning to the controller. The controller sends the flow tables to cache Avatar in this switch so as to reduce users' waiting time for Avatar.

\subsection{Specific Example}
For example, if our objective is to achieve minimal network flow by addressing the questions of how each content router caches {\color{black}content} in the network and how to allocate limited cache resource among content routers, the total problem can be formulated as follows:

To begin, we define the following quantities that can be pre-computed or defined.
\begin{enumerate}
\item \emph{$V$}: the set of network nodes, which is $V=\{V_{1}, V_{2},...,V_{N}\}$, indexed by $i,j$.
\item \emph{$O$}: the set of content objects, which is $O=\{O_{1},O_{2},...,O_{M}\}$, indexed by $O_{k}$.
\item \emph{$C_{sum}$}: it is the sum of cache size of all caches in the system.
\item \emph{$S_{k}$}: it is the size of the content object $O_{k}$.
\item \emph{$d_{ijk}$}: it is the hop distance by node $i$ to request content object $O_{k}$ from node $j$.
\item \emph{$q_{i}^{k}$}: it is the request rate for content object $O_{k}$ at node $i$.
\end{enumerate}

We introduce the following variables:
\begin{enumerate}
\item \emph{$y_{ijk}$}: $y_{ijk}$ takes the value of $1$ if node $i$ downloads a copy of content object $O_{k}$ from node $j$, and $0$ otherwise.
\item \emph{$x_{ik}$}: $x_{ik}$ takes the value of $1$ if node $i$ caches a copy of content object $O_{k}$, and $0$ otherwise.
\item \emph{$C_i^t$}: the cache size of node $i$ at $t$ time.
\end{enumerate}

Then, the formulation is:
\begin{equation}
\begin{aligned}
\textbf{min:}\sum_{i=1}^N\sum_{k=1}^Mq_{i}^k\sum_{j=1}^N
d_{ijk}S_{k}y_{ijk}
\end{aligned}
\end{equation}

s.t.:
\begin{equation}
\begin{aligned}
&\sum_{j=1}^Ny_{ijk}=1, \forall i,k \\
&y_{ijk} \leq x_{ik}, \forall i,j,k \\
&\sum_{k=1}^Mx_{ik}S_{k}\leq C_{i}^t, \forall i \\
&\sum_{i=1}^NC_{i}^t=C_{sum}, \forall t
\end{aligned}
\end{equation}

The first constraint specifies that each content router $i$ downloads object $k$ from only one content router. The second constraint specifies that content router $i$ downloads object $k$ from content content router $j$ only when content object is located there. Then, the third constraint $\sum_{k=1}^Mx_{ik}S_{k}\leq C_{i}^t, \forall i$ specifies that the total size of content objects located in content router $i$ can't exceed the current maximum cache size. Finally, the constraint $\sum_{i=1}^NC_{i}^t=C_{sum}, \forall t$ specifies that the sum cache size of all content routers remains the same.

Leveraging the big data algorithm and optimization algorithm in the algorithm set, we can find the optimal solution to this problem.  Finally, the  big data processing platform exports optimal solutions to the controller. By these optimal solutions, the controller manages traffic, content allocation and cache distribution.

With that consideration, we implement the experimental simulation. Based on the Open Shortest Path First (OSPFN) routing protocol, the performance of the proposed architecture is evaluated in the ndnSIM 1.0 simulator and Hadoop.

In the ndnSIM, we simulate the request, reply, distribution of {\color{black}content}, data encapsulation and content forwarding. After that, we obtain the necessary user data from ndnSIM and send to Hadoop. With the help of big data analysis in Hadoop, we can extract knowledge out of the large volumes of data from ndnSIM and get the decision about how to allocate cache resource, distribute {\color{black}content}. Finally, resend
those decision to the ndnSIM to reallocate cache resource and redistribute {\color{black}content}.

\subsubsection{Simulation Settings}
\begin{itemize}
    \item Network Topologies: The simulation is carried out in the Power-Law topology including 64 content routers.
    \item Input Data: There is 200 different {\color{black}content} in the simulation. We assume each object has the same size and the content popularity follows the Zipf distribution, in which the skewness factor $\alpha =0.8$ \cite{choi_-network_2012}.
    \item Cache Size: We abstract the cache size for each content router as a proportion that the cache size is defined as the relative size to the total amount of different {\color{black}content} in the network. We evaluate the network performances for each caching scheme when the cache memory size varies from $1\%$ to $10\%$.
\end{itemize}

\subsubsection{Performance Evaluation Results}
\begin{figure}[tp]
\centering
\includegraphics[width=8cm]{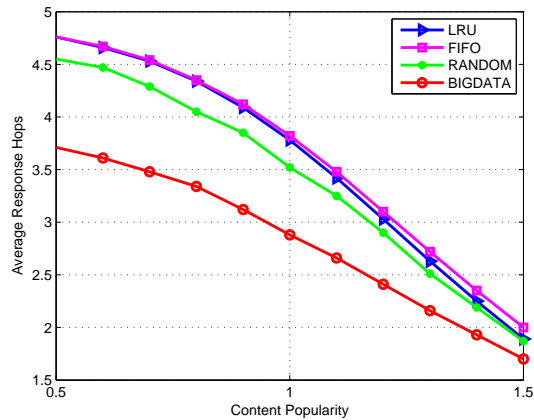}
\caption{Average response hops versus content popularity.}
\label{fig: fig5}
\end{figure}

Fig.~\ref{fig: fig5} shows the average response hops of different cache policies when the content popularity varies. From Fig.~\ref{fig: fig5}, we can observe that the content popularity has some effects on the delay of each cache scheme. When the content popularity increases, the number of popular {\color{black}content} changes in the same way, which improves the average response hops of each solution. Besides, it also makes the gap among each scheme smaller, because the influence of different cache policies with a higher content popularity is weakened. However, the proposed solution always has a better performance, because the cache decision is made based on the global knowledge, which makes the network better adapt to the change of the network.

\begin{figure}[tp]
\centering
\includegraphics[width=8cm]{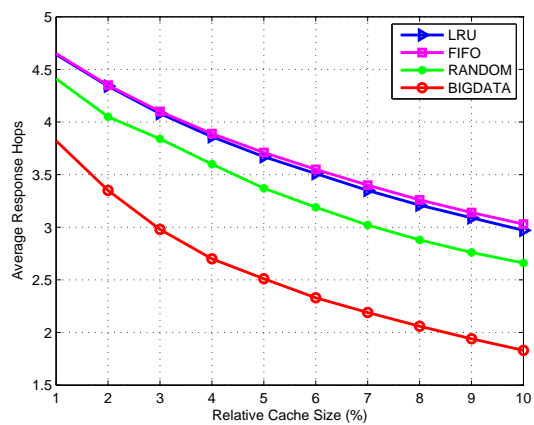}
\caption{Average response hops versus cache size.}
\label{fig: fig6}
\end{figure}

Fig.~\ref{fig: fig6} shows the average response hops of different cache policies when cache size varies. From Fig.~\ref{fig: fig5}, we can observe that the cache has some effects on the delay of each cache scheme. When the cache size increases, more popular {\color{black}content is} cached, which makes the average response hops of each solution smaller. Besides, it also makes the gap among each scheme larger, because the influence of different cache policies with a larger memory becomes more obvious. However, the proposed solution always has a better performance, because the cache decision is made based on the global knowledge, which makes the network cache optimal {\color{black}content} in the extra buffer.

\section{Conclusions and Future Work\label{sect: con}}
In this paper, we have proposed a novel data-driven network architecture, in which in-network caching is added in the infrastructure layer of SDN and a  big data analysis module is added in the control layer of SDN. In particular, each SDN switch has the caching capability to facilitate efficient content distribution. With the help of centralized SDN controller, the system with big data analysis can be aware of the information about users, {\color{black}content} and network to realize optimal resource allocation, efficient content distribution and flexible network configuration. Simulation results were presented to show that this novel network architecture can efficiently improve the network performance compared to the tradition schemes. Future work is in progress to considering cloud/fog computing in the proposed network architecture.

\section*{Acknowledgment}
The project is supported by the Key Program of the National Natural Science Foundation of China (Grant No 61431008 and 61471056).

\bibliographystyle{IEEEtran}
\bibliography{GLOBECOM}
%% insert where needed to balance the two columns on the last page with
%%\newpage
%\end{IEEEbiography}
%dddd
%\end{IEEEbiography}
%%\begin{IEEEbiographynophoto}{Jane Doe}
%%Biography text here.
%%\end{IEEEbiographynophoto}
%
%% You can push biographies down or up by placing
%% a \vfill before or after them. The appropriate
%% use of \vfill depends on what kind of text is
%% on the last page and whether or not the columns
%% are being equalized.
%
%%\vfill
%
%% Can be used to pull up biographies so that the bottom of the last one
%% is flush with the other column.
%%\enlargethispage{-5in}
%
%\bibliographystyle{IEEEtran}
%\bibliography{GLOBECOM}

%% that's all folks
\end{document}